\documentstyle[11pt,aaspp4,epsfig]{article}

\def\simg{\mathrel{\hbox{\rlap{\lower.55ex \hbox {$\sim$}}
                   \kern-.3em \raise.4ex \hbox{$>$}}}}
\def\siml{\mathrel{\hbox{\rlap{\lower.55ex \hbox {$\sim$}}
                   \kern-.3em \raise.4ex \hbox{$<$}}}}
\def\Mesz{M\'esz\'aros~}
\def\Pacz{Paczy\'nski~}
\def\beq{\begin{equation}}
\def\enq{\end{equation}}
\def\bea{\begin{eqnarray}}
\def\ena{\end{eqnarray}}

\def\bec{\begin{center}}
\def\enc{\end{center}}
\def\etal{{\it et al.}}
\def\ergs{\hbox{erg s$^{-1}$}}
\def\ergcmsqs{\hbox{erg cm$^{-2}$ s$^{-1}$}}
\def\cm2si{\hbox{cm$^{-2}$s$^{-1}$}}

\def\gcmcui{{\rm g~cm}^{-3}}
\def\Liso{L_{iso}}
\def\L52{L_{52}}
\def\gammad{{\bar \gamma}}
\def\eldif{\ell_{diff}}

\def\tv3{t_{v-3}}

\begin{document}

\title{ Collapsar Uncorking and Jet Eruption in GRB}


\author{E. Waxman$^{1}$ and P. \Mesz$^{2}$}
\smallskip\noindent
\bec
$^1${Weizmann Institute of Science,  Rehovot 76100, Israel} \\
\smallskip\noindent
$^2${Pennsylvania State University, University Park, PA 16803}
\enc

\bec ApJ, subm. :~{6/18/02} --  Revised subm. :~{9/25/02 } \enc

\begin{abstract}
We show that the collapsar model of gamma-ray bursts results in a
series of successive shocks and rarefaction waves propagating in
the ``cork" of stellar material being pushed ahead of the jet, as
it emerges from the massive stellar progenitor. Our results are
derived from analytical calculations assuming a hot,
ultrarelativistic 1-D flow with an initial Lorentz factor
$\Gamma_j\sim 100$. The shocks result in a series of
characteristic, increasingly shorter and harder thermal X-ray
pulses, as well as a non-thermal $\gamma$-ray pulse, which precede
the usual non-thermal MeV $\gamma$-rays. We consider jets escaping
from both compact (CO or He dwarf) and blue supergiant stellar
progenitors. The period, fluence and hardness of the pulses serves
as a diagnostic for the size and density of the outer envelope of
the progenitor star.

\end{abstract}

\section{Introduction}

The non-thermal radiation from shocks in a relativistic fireball
jet after emerging from a collapsing massive stellar system
(collapsar) is the leading theoretical explanation for the family
of ``long" gamma-ray bursts (GRB) with $\gamma$-ray durations
longer than about $t_b\sim 10$ s (Woosley 1993, \Pacz 1998; see
\Mesz 2002 for a review). The jet is thought to arise from a brief
accretion episode of a rotating debris torus around a central
black hole formed as the stellar core collapses. The nature and
history of the massive star, however, is so far largely a matter
of hypothesis and calculation, and alternative stellar progenitor
systems are conceivable, both for the long and especially for the
shorter bursts.

The duration of the TeV neutrino burst expected from shocks in the
jet before it emerges is a possible diagnostic for the
pre-emergence jet history and for the dimensions or column density
of the progenitor stellar system (\Mesz \& Waxman 2001). Previous
analytical work on shock and jet emergence has been done by
Matzner \& McKee 1999, \Mesz \& Rees 2001, Matzner 2002, and
numerical calculations of relativistic jets have been done by
Marti, M\"uller, Font, Iba\~nez \& Marquina, 1997; Aloy, M\"uller,
Iba\~nez, Marti \& MacFadyen, 2000; and Zhang, Woosley \&
McFadyen, 2002.
Here we concentrate on possible photon signatures of the jet as it
emerges, which precedes the usual gamma-ray emission and the
subsequent longer wavelength afterglow. This has been considered
by Ramirez-Ruiz, McFadyen \& Lazzati (2001), who infer an X-ray
precursor to the burst from the portion of the stellar envelope
shocked by the jet. Here we investigate in some detail, using
analytical 1-D methods in the ultrarelativistic limit, the jet
emergence process and the shock heating and expansion of the plug
of stellar material propelled by it, resembling an ejected cork.

In the initial stage of ejection the cork is optically very thick
to scattering, and it experiences successive shocks and
rarefactions as the optical depth decreases during its overall
expansion. This leads to an X-ray photon precursor to the usual
GRB emission, consisting of a pattern with an initial rise and
decay followed by a brighter and harder peak, possibly repeated a
few times with decreasing period.
The characteristic timescale of the initial decay and the ratio of its
amplitude to that of the peak differ substantially depending on the
density and extent of the envelope, thus providing a potential diagnostic
for the pre-burst stellar configuration. These precursor patterns should
be detectable with instruments to be flown in the next few years.

\section{Jet propagation}
\label{sec:jet-prop}

\begin{figure}[htb]
\centering \epsfig{figure=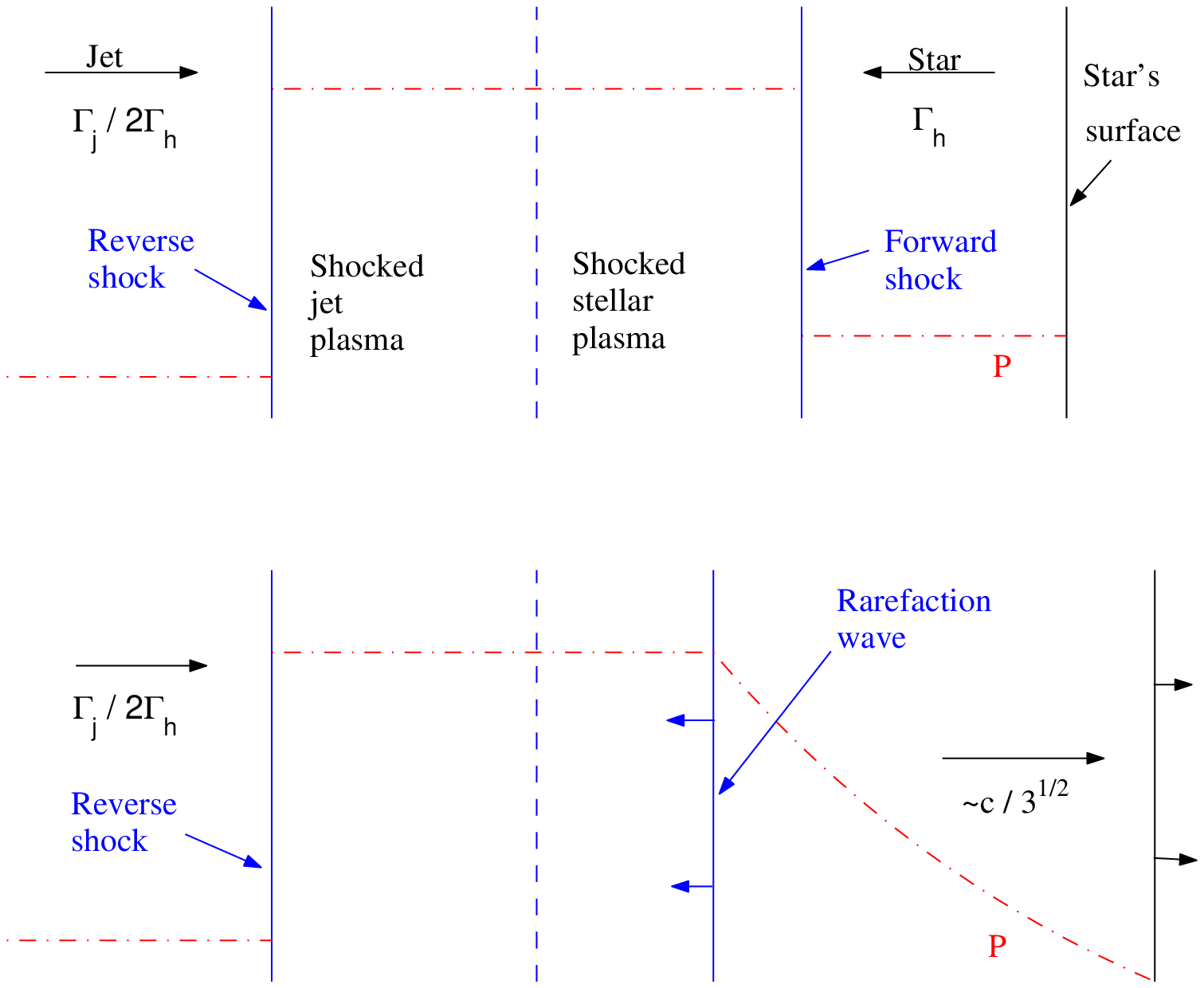} \caption{The top panel
presents a schematic description of the double shock structure at
the head of the jet, as seen in the frame where the shocked plasma
is at rest, during jet propagation within the star. The dot-dashed
line shows the pressure distribution. The bottom panel presents a
schematic description of the flow after forward-shock break-out. A
rarefaction wave propagates backward into the shocked plasma, and
the plasma ahead of this wave expands with velocity comparable to
the speed of sound. When the rarefaction wave reaches the backward
shock, a new double-shock structure similar to that shown on the
top panel is formed, propagating through the expanding shocked
plasma.}
   \label{fig:shocks}
\end{figure}

We assume a jet with isotropic equivalent luminosity $\Liso \simg
10^{52}\L52 ~\ergs$ and Lorentz factor $\Gamma_j\sim 10^2$. As the
jet advances through the star, it drives forward a bow shock into
the star. The jet is capped by a "cork" composed of a termination
shock advancing into the star, and a reverse shock moving back
into the jet, where the tenuous and highly relativistic jet gas is
suddenly decelerated and compressed. Both the shocked jet plasma
and the shocked stellar plasma advance inside the star with a jet
head Lorentz factor $\Gamma_h \ll \Gamma_j$, corresponding to a
velocity $\beta_h$. Figure \ref{fig:shocks} presents a schematic
description of the multiple shock structure. In the rest frame of
the shocked, decelerated plasma, the fast jet moves with a Lorentz
factor $\sim \Gamma_j /2 \Gamma_h$. During propagation in the He
core, the high density ahead of the jet, $\rho\sim 1 ~\gcmcui$,
implies deceleration of the jet plasma to a sub-relativistic
velocity. However, beyond the C/O or He core edge at $r\simg
10^{10}- 10^{11}$ cm, there is a sharp drop in the density, e.g.
the density of an H envelope, if present, is typically $\rho \sim
10^{-5}-10^{-7}\gcmcui$, and the jet can accelerate to
relativistic velocity. We have used Matzner \& McKee's
parametrization of the density drop near the edge of the star,
$r=R_*$, \beq \rho=\rho_*\left(\frac{R_*}{r}-1\right)^n,
\label{eq:edge} \enq where $n=3\,(3/2)$ for radiative (convective)
envelope. Eq. (\ref{eq:edge}) holds for a polytropic plasma at
radii beyond which the envelope contribution to the gravitational
potential is small. We consider three progenitor examples: i) the
edge of the C/O progenitor model of SN1998bw (Woosley, Eastman \&
Schmidt 1999), approximately described by $\rho_*=5\times10^2{\rm
g/cm}^3$, $R_*=10^{10}$~cm and $n=3$; ii) the outer part of the He
mantle of the pre-supernova model, evolved by Woosley, Langer \&
Weaver (1993) from a $35M_\odot$ main sequence star, approximately
described by Eq. (\ref{eq:edge}) with $\rho_*=2{\rm g/cm}^3$,
$R_*=10^{11}$~cm and $n=3$; and iii) the envelope of the blue
supergiant (BSG) progenitor models of SN1987A (Shigeyama \& Nomoto
1990,Arnett 1991), approximately described by
$\rho_*=3\times10^{-5}{\rm g/cm}^3$, $R_*=3\times10^{12}$~cm and
$n=3$. Taking in Eq.(\ref{eq:edge}) a radius $r\simeq 0.9 R_*$ we
take as approximate reference quantities the values $\rho\sim
[1,10^{-3}, 10^{-7}]\gcmcui$ at $r\sim [10^{10},  10^{11},
10^{12.5}]$ cm for a   C/O core, He core and BSG star H envelope
respectively.

The pressure behind the forward shock is given, for $\Gamma_h \gg 1$,
by $P_h=(4/3)\Gamma_h^2\rho c^2$, and for $\beta_h\ll 1$, by
$P_h= [(\gammad +1)/2]\rho\beta_h^2 c^2$, where $\gammad=4/3$ is the
adiabatic index of the plasma (radiation dominated).
Equating $P_h$ with the pressure behind the reverse shock,
$P_r=(4/3)(\Gamma_j/2\Gamma_h)^2 n_j m_p c^2$ for $\Gamma_h \gg 1$
and $P_r=(4/3)\Gamma_j^2 n_j m_p c^2$ for $\beta_h\ll 1$, where
the jet proper proton density $n_j=\Liso/(4\pi r^2\Gamma_j^2 m_p
c^3)$ (For a discussion of relativistic flows and shock jump
conditions, see e.g., Begelman, Blandford \& Rees 1984). The
extreme relativistic limit gives (\Mesz \& Waxman 2001) \beq
\Gamma_h \sim \cases{
  1.6{\L52^{1/4} / r_{12.5}^{1/2}\rho_{-7}^{1/4}} & H;\cr
  0.9 {\L52^{1/4} / r_{11}^{1/2}\rho_{-3}^{1/4}} & He \cr}
\label{eq:Gammah}
\enq
for the H-envelope of a BSG (H) or a He star, while the
non-relativistic limit gives
\beq
\beta_h \sim \cases{
  1.6{\L52^{1/2}/ r_{11}\rho_{-3}^{1/2}} & He; \cr
  0.6{\L52^{1/2}/ r_{10}\rho_{0}^{1/2}}  & CO \cr}
\label{eq:betah} \enq for the He or CO cores, where more exactly
$\beta_h\leq 1$ and $\Gamma_h \geq 1$ would hold;
$\Liso=10^{52}\L52\ergs$ is the jet isotropic equivalent
luminosity, $r=10^x r_x$ is the shock radius, and $\rho=
10^x\rho_x\gcmcui$ is the corresponding density at that radius.
These values are for the main part of the cork, including the
shocked jet and  the bulk of the shocked stellar material. Thus,
for $r\sim 0.9R_*$ in all cases the jet head velocity is at least
marginally relativistic, and this approximation gets better as
$r\to R_*$. This implies that, even though the jet injected at the
base of the jet is assumed to be ultrarelativistic, $\Gamma_j=10^2
\Gamma_2 \gg 1$, the jet head advances sub-relativistically for
most of its initial crossing of the stellar interior. This is due
to the initial inertia of the dense stellar material which needs
to be pushed ahead and aside (e.g. \Mesz \& Rees, 200 for an
analytical discussion). A similar relativistic injection and
subrelativistic jet head advance occurs in AGN jets encountering
the intergalactic medium, e.g. Begelman \etal 1984. The jet
crossing time out to the He (or CO) core edge is of order
$t_{core}\siml 30 r_{11}$ s for a jet which has reached $v_h\sim
c$ near the core edge, suggesting that compact ($r\siml 10^{11}$
cm) stars such as He stars or pre-WR stars are likely GRB
progenitors (see also MacFadyen \etal~2001 and Matzner 2002).

However, if there is an H-envelope beyond the C/O/He core, where
the density drops steeply by $10^{-5}-10^{-7}$, momentum
conservation indicates that $\Gamma_h$ for the entire cork
(equation [{\ref{eq:Gammah}]) can become significantly larger at
radii just beyond $r_{core}=10^{11}R_{11}$ cm. This jump in
$\Gamma_h$ is absent in Matzner's (2002) treatment, which
approximates with an unvarying $\rho\propto r^{-2}$ dependence
the entire density profile on either side and across the He-H
transition. A density drop steeper than $r^{-2}$ is however
expected at this core-envelope transition for radiative envelopes
(Eq.[\ref{eq:edge}]), and is a distinctive feature of the
numerical pre-collapse models employed above. Using equation
(\ref{eq:Gammah}) for the Lorentz factor of the entire cork in an
H-envelope, the additional crossing time can be much much smaller
than the core crossing time already incurred, $t_{H}\sim
r_H/2c\Gamma_h^2 \ll t_{core}$ (\Mesz \& Rees, 2001), as long as
the envelope is not too extended, $r_H \siml \hbox{few}\times
10^{12}$ cm, e.g. for blue supergiants (BSG). This is independent
of whether a small fraction of the outermost cork mass is
accelerated along the density gradient to larger Lorentz factors
(Matzner \& McKee 1999), as discussed in the next section. The
only other restriction besides the envelope size is that the
envelope mass should be smaller than that inside the stellar core,
which is generally true for BSGs. Thus, the jet ``powering" time
$t_j$ needed to produce an emerging jet (and hence a successful
GRB) is limited mainly by the core crossing time $t_{core}\sim 30$
s, which is similar both for envelope-stripped CO or He stars and
for stars as extended as blue supergiants with H-envelopes of $r_H
\siml \hbox{few}\times 10^{12}$ cm.

The cork is composed of shocked jet and stellar plasma. The
comoving width of the shocked jet plasma is $\Delta_j=0.2\theta r$
(\Mesz \& Waxman 2001), where $\theta$ is the asymptotic jet
opening angle. The comoving width of the shocked stellar plasma is
obtained by balancing the mass flux across the bow shock,
$\approx\pi(\theta r^2)(\beta_h c+v'_s)(\Gamma_h\rho)$ where
$v'_s$ is the shock velocity in the cork frame, with the
tangential flux of particles leaving the cylinder of (cork frame)
height $\Delta_s$ and radius $\theta r$ of shocked plasma,
$\approx 2\pi\theta r \Delta_s c_s \rho'$ , where $c_s$ and
$\rho'$ are the post-shock speed of sound and density
respectively. Using $c_s=c/\sqrt{3}$ and $\rho'=4\Gamma_h\rho$ for
the relativistic ($\Gamma_h \gg 1$) case, which from now on we
take as approximately valid for all cases considered, we obtain
$\Delta_0 \simeq 0.2 \theta r$. The Thomson optical depth of the
shocked stellar plasma is thus \beq \tau_{T,s}\simeq \cases{
 1.5\times 10^4 \theta_{-1}\L52^{1/4} r^{1/2}_{12.5}\rho_{-7}^{3/4},& H ;\cr
 2.7\times 10^6 \theta_{-1}\L52^{1/4} r^{1/2}_{11}\rho_{-3}^{3/4},& He ;\cr
 1.5\times 10^8 \theta_{-1}\L52^{1/4} r^{1/2}_{10}\rho_{0}^{3/4},& CO . \cr}
\label{eq:tauTs} \enq
Here, $\theta=0.1\theta_{-1}$. The cork
temperature is \beq T_r\simeq \cases{1.6 \L52^{1/8}
r^{-1/4}_{12.5} \rho_{-7}^{1/8},& H ; \cr
                 12  \L52^{1/8} r^{-1/4}_{11} \rho_{-3}^{1/8},& He ;\cr
                 50  \L52^{1/8} r^{-1/4}_{10} \rho_{0}^{1/8},& CO .\cr }
            \quad{\rm keV}.
\label{eq:Tr}
\enq
The corresponding (cork frame) energy carried by the cork is thus
\beq
E_0\simeq \cases{
  3.9\times 10^{48}\L52^{3/4}\theta_{-1}^3 r_{12.5}^2, & H;\cr
  4.3\times 10^{45}\L52^{3/4}\theta_{-1}^3 r_{11}^2, & He;\cr
  4.3\times 10^{43}\L52^{3/4}\theta_{-1}^3 r_{10}^2, & CO.\cr}
\quad{\rm erg}. \label{eq:E0}
\enq

\section{Jet emergence: X-ray precursor}
\label{sec:emergence}

\subsection{Dynamics}
\label{sec:emergence-dynamics}

As the forward shock approaches the stellar edge it accelerates to
relativistic velocity. At some finite depth $d=R-r$ it becomes
causally disconnected from the backward shock. This occurs at the
point where during a sound crossing time of the cork, $\delta
t\approx\Gamma_h 0.4\theta r/c_s$, the forward shock accelerates
significantly. Since $\Gamma_h\propto\rho^{-1/4}$ (as long as the
forward shock is causally connected to the backward one), the
fractional change in the shock Lorentz factor is
$\delta\Gamma/\Gamma\approx\delta\rho/4\rho=c\delta t
d\ln\rho/dr\approx 0.4\Gamma_h(c/c_s)\theta r (n/4)/(R_*-r)$.
Thus, the forward shock becomes disconnected at $d\equiv R_*-r
\approx\Gamma_h(d)\theta r$, where $\delta\Gamma/\Gamma\approx 1$
(here $\Gamma_h(d)$ is the main cork Lorentz factor at depth $d$).
Denoting by $\Gamma_*$ the value of $\Gamma_h$ given by Eq.
(\ref{eq:Gammah}) for $r=R_*/2$ assuming the density profile of
Eq. (\ref{eq:edge}), we find
$d/R_*\approx(\Gamma_*\theta)^{4/(n+4)}$ and $\Gamma_h(d)\approx
\Gamma_*^{4/(n+4)}\theta^{-n/(n+4)}$. Assuming
$\theta\approx1/15$, the value inferred by Frail et al. (2001), we
find that the cork is mildly relativistic at the onset of
``runaway" for all the progenitors (He, BSG, C/O) considered
above. The dependence of $\Gamma_h(d)$ on $\theta$ is not very
strong, $\Gamma_h(d)\propto\rho(d)^{-1/4}\propto(d/R)^{-n/4}
\propto\theta^{-n/(n+4)}$. The stellar densities at ``runaway"
onset are $\sim1{\rm g/cm}^3$, $\sim10^{-3}{\rm g/cm}^3$ and
$\sim10^{-7}{\rm g/cm}^3$ for the C/O, He and BSG progenitors
respectively.

As the forward shock reaches depth $d$, it accelerates and breaks
through the stellar edge before the backward shock reacts to the
rapid acceleration. During jet propagation in the envelope, the
shocked cork plasma is pushed sideways, away from the jet path, on
a time scale $\Gamma_h\theta r/c_s$. Since the forward shock
``runaway", from depth $d$ outwards, takes place on a similar time
scale, we may assume that the cork is ``frozen" during this time.
At break-out we will therefore have a cork of thickness $0.4\theta
R_*$ with density, temperature and velocity given by the equations
of \S\ref{sec:jet-prop} with $r=R_*$ and density $\rho(R_*-d)$.
Ahead of this cork, there will be a faster moving shell of plasma,
accelerated during shock ``runaway" (similar to that considered in
spherical supernova shocks by Matzner \& McKee 1999). The column
density of the faster plasma is smaller than that of the cork by a
factor $\approx 1/(n+1)$.

During the shock runaway, the pressure behind the forward shock is
smaller than that behind the backward shock (since the two are
causally disconnected). The forward shock Lorentz factor is
therefore smaller than that given by  Eq. (\ref{eq:Gammah}),
using the density ahead of the forward shock: $\Gamma(x)\approx
(x/d)^{-n/4}$ where $x$ is the depth. The optical depth at depth
$x$ is $\tau(x)\approx[\tau_{T,s}/(n+1)](x/d)^{n+1}$, where
$\tau_{T,s}$ is the cork optical depth, implying
$\Gamma(x)\approx[\tau_{T,s}/(n+1)\tau(x)]^{n/4(n+1)}$. Since the
shocks are radiation dominated, the maximum Lorentz factor to
which plasma is accelerated is smaller than that given by this
expression for $\tau(x)=1$. Using Eq. (\ref{eq:tauTs}), we find
that the maximum Lorentz factor is of order 10 for all progenitor
types mentioned above.

Whether or not a fast, $\Gamma\sim10$, shell is ejected ahead of
the cork is unclear. As the forward shock accelerates and
becomes disconnected from the driving jet "piston", it is likely
to become unstable (as commonly found in the non-relativistic
case, e.g. Sari, Waxman \& Shvarts 2001). Moreover, if the
progenitor plasma is magnetized, the magnetic field may also
prevent large $\Gamma$ gradients. We therefore consider in the
following sub-section X-ray emission from the slower, main cork
plasma. We briefly comment here on how a fast shell ahead of the
cork may affect this X-ray emission.

Photons emitted by the cork would overtake the fast shell, spend
some time diffusing through it, and then escape. Let us first
consider the time delay introduced to photon arrival times seen by
a distant observer. We show below that the fast shell does not
change its radius significantly during the photon diffusion through
it. The time delay may therefore be decomposed into two separate
contributions: A geometric time delay due to a change in the
photon propagation direction, which is assumed to take place
instantaneously at the shell overtaking radius, and a diffusion
time delay due to the fact that the effective velocity of the
photon during the time it spends in the shell is smaller than $c$
(by $c/2\Gamma(x)^2$).

As shown below, the time scale of photon emission from the cork is
$t_0\sim\theta R_*/c$. Photons emitted radially from the cork at
time $t$ following break-out will overtake a fast shell with
$\Gamma(x)$ at time $t_1\sim2\Gamma(x)^2t$. For $t\gtrsim t_0$, a
fast shell with $\Gamma(x)>\theta^{-1/2}$ expands significantly by
$t_1$, $c t_1/R_*\sim \Gamma(x)^2\theta$. In this case, the spread
in overtaking times between photons emitted radially and
no-radially from the cork, $\sim\theta^2 R_*/c$, is small compared
to $t_0$, and we may consider geometric time delay of radial
photons only. For an observer along the jet axis, the geometric
time delay between two photons emitted at time $t$ and scattered
towards the observer from different points in the fast shell, one
from a point along the line of sight and the other from the edge
of the shell, propagating at angle $\theta$ with respect to the
line of sight (we have shown above that $\Gamma(x)<10 \lesssim
1/\theta$ so that such scattering is possible), is $\approx
0.5\theta^2 t_1\approx [\Gamma(x)\theta]^2 t$. Since $\Gamma(x)
\theta<1$, the time spread introduced by this geometric time delay
would not affect significantly the shape of the pulse emitted by
the cork plasma.

Let us now consider the diffusion time delay. For $t\gtrsim t_0$,
the optical depth of the shell at $t_1$ is $\tau_1(x)\approx
\tau(x)(c t_1/R_*)^{-2}$. Since the proper thickness of the fast
shell is $x/\Gamma(x)$ (due to the shock compression), the photon
diffusion time through the shell is $\approx\tau_1 x/\Gamma(x) c$
in the shell frame, corresponding to $\delta t_{diff}\approx\tau_1
x/c$ in the star frame. The delay in photon propagation (compared
to free propagation) is $\delta t_{prop}\approx\delta
t_{diff}/2\Gamma(x)^2$. For $t>t_0$ we thus find $\delta
t_{prop}/t_0\sim[\tau_{T,s}/(n+1)\theta^2](t/t_0)^{-2}\Gamma(x)^{-2(5n+4)/n}$.
For $n=3$, the diffusion time delay is small, compared to $t_0$,
for $\Gamma(x)>[\tau_{T,s}/(n+1)\theta^2]^{3/38}$. Using Eq.
(\ref{eq:tauTs}), we conclude that this time delay is small for
$\Gamma(x)>$~a few. The fractional change in the shell radius
during photon diffusion, $\sim\delta t_{diff}/2\Gamma(x)^2
t\approx\delta t_{prop}/t<\delta t_{prop}/t_0$, is also small,
justifying our assumption that the shell radius is approximately
constant over a photon diffusion time.

The arguments given above demonstrate that the fast shell would
not affect significantly the pulse shape. Let us now consider its
effect on the photon spectrum. The comoving post-shock temperature
decreases towards the stellar edge,
$T(x)^4\propto\Gamma(x)^2\rho(x)\propto\rho(x)^{1/2}$, and the
apparent temperature $\Gamma(x)T\propto\Gamma(x)^{1/2}$. The
apparent temperature of the fast shell is not much larger than
that of the cork, any expansion of the shell further reduces its
temperature, and its energy content is much smaller than that of
the main cork. An inverse Compton contribution is also found
to be much less energetic than other components
(\S \ref{sec:non-thermal}). Thus, we expect the fast shell to have
little effect on the spectrum as well.

\subsection{X-ray flash at the first shock emergence}
\label{sec:1stflash}

We consider now the X-ray emission from the cork break-out. We
assume that the cork expands at its speed of sound at the cork
break-out time, $c_s \simeq c/\sqrt{3}=0.58c$ in the relativistic
case. This is the sound speed for a radiation-dominated gas,
rarefaction fronts typically expanding at the speed of sound. As
discussed before, at cork break-out the shock is at least mildly
relativistic in the cases of interest here, so we use this limit
from now on).
For simplicity we replace the flattened cork (width $=0.4\theta
r<{\rm diameter}=2\theta r$) with a sphere, and denote its radius
by $\Delta_0$ $=2\theta r/3$, which has the same volume. (The
spherical expansion description in the comoving frame is a rough
but convenient approximation, which becomes increasingly
justifiable as the cork expands beyond one e-folding of its
original dimensions at the same speed in all directions.) The
cork's radius increases (in its comoving frame) as
$\Delta=\Delta_0 +c_s t$. We therefore define the expansion factor
\begin{equation}
x\equiv  1+c_s t/\Delta_0,
\label{eq:x}
\end{equation}
so that $\Delta=\Delta_0 x$, $\rho=\rho_0 x^{-3}$, $T_r=T_{r,0} x^{-1}$ and
$\tau_T=\tau_{T,0} x^{-2}$. Subscripts $0$ denote values at cork break-out
time. Since the optical depth is very large, photons diffuse out of a thin
skin depth at the edge of the cork. The width of the skin-depth layer
out of which
photons have escaped is $\eldif \simeq \sqrt{c t \Delta /3 \tau_T}$, where
$\Delta /\tau_T$ is the photon mean free path. We therefore have
$\eldif/\Delta\simeq \sqrt{c t /3 \Delta \tau_T}=
(x/\tau_{T,0})^{1/2}(t/t_o)^{1/2}$,
where $t_0\equiv3\Delta_0/c$.
The amount of energy radiated up to time $t$ is
\beq
E_{rad}\simeq 4\pi \Delta^3 {\eldif \over \Delta} a T^4  =
  {3 E_0 \over \sqrt{\tau_{T,0}}} \left({t \over x t_0}\right)^{1/2}~,
\label{eq:Erad}
\enq
and hence the comoving luminosity is
\beq
L_{rad}\simeq
 { 3E_0 \over 2 t_0 \sqrt{\tau_{T,0}}} x^{-3/2} \left({t\over t_0} \right)^{-1/2}~.
\label{eq:Lrad}
\enq
For times $t\ll t_{exp} \equiv\Delta_0/c_s=(c/3c_s)t_0$,
the expansion is not important, $x\simeq 1$, hence $L\propto t^{-1/2}$ and
$T_r\simeq T_{r,0}$. At later times, $x\gg 1$, $L\propto t^{-2}$ and
$T_r  \propto t^{-1}$. We note that the speed of sound in the shocked
jet plasma is always (even for $\beta_h\ll1$) relativistic, so that
the rarefaction wave crosses the shocked jet plasma much faster than it
crosses the shocked stellar plasma. Since the internal energy
stored in the shocked jet and stellar plasma are similar (from the
momentum and energy conservation across the shock implied by the
shock jump conditions), the expansion of the shocked jet plasma
will accelerate the expansion of the shocked stellar plasma to a
velocity larger by a factor $\sim\sqrt{2}$. Thus, for arbitrary
$\beta_h$, whether sub-relativistic or relativistic, we can
approximate $t_{exp}\approx t_0/2\beta_h$. The total energy
radiated is thus \beq E_{tot}\simeq
  {3 E_0 \over \sqrt{2\beta_h\tau_{T,0}}}.
\label{eq:Etot} \enq
One should keep in mind the potential
limitations of the 1-D approximation used here, since in 2-D
numerical calculations (e.g. Aloy \etal, 2000, Zhang \etal, 2002)
the emergence of the reverse shock and rarefaction wave indicates
the formation of a conical shape due to lateral expansion. For the
parameters we used here, the 1-D approximation provides a useful
description as long as the jet has not moved beyond much beyond
$\sim R_\ast \theta$.

For a point on the source-observer line of sight, the
characteristic break-out timescale $t_{exp}$ is shortened to
$t_{exp}^{obs}=t_{exp}/2\Gamma_h= \Delta_0/2\Gamma_h c_s\approx
0.4\theta r/2\Gamma_h c_s$. The angular smearing of the light
curve occurs on a timescale $d t_{ang}^{obs} \simeq \theta^2
r/2c$, for which $dt_{ang}^{obs}/t_{exp}^{obs}= 5\Gamma_h \theta
c_s / 2 c < 5\theta \Gamma_h /2\sqrt{3}$. Since $\theta \ll 1$ and
$\Gamma_h$ is at most a few, we have $dt_{ang}^{obs}/t_{exp}^{obs}
\ll 1$ and at a distance $d$ the flux $f \sim L/d^2$ of the
observed pulse shape follows Eq. (\ref{eq:Lrad}): $f\propto
t^{-1/2}$ for $t<t_{exp}^{obs}$, \beq t_{exp}^{obs}= \cases{
 2.2 \L52^{-1/4}\theta_{-1}r_{12.5}^{3/2}\rho_{-7}^{1/4}, &H ;\cr
 1.4\times 10^{-1}\L52^{-1/4}\theta_{-1}r_{11}^{3/2}\rho_{-3}^{1/4},& He ;\cr
 2.4\times 10^{-2}\L52^{-1/4}\theta_{-1}r_{10}^{3/2}\rho_{0}^{1/4}, &CO ,\cr}
 \quad{\rm s}, \label{eq:texp_obs}
\enq and $f\propto t^{-2}$ at later times (here $d$ is distance of
the source). The flux $f_0$ at $t=t_{exp}^{obs}$ is approximately
given by $f_0=\Gamma_{h}
(3E_0/\sqrt{2\tau_{T,s}})/(2t_{exp}^{obs}) \pi(d/\Gamma_h)^2$ in
the relativistic approximation,
\beq
f_0= \cases{
2.2\times 10^{-9}\L52^{11/8}\theta_{-1}^{3/2}/r_{12.5}^{5/4}\rho_{-7}^{7/8}d_{28}^2 ,&H ;\cr
4.9\times 10^{-11}\L52^{11/8}\theta_{-1}^{3/2}/r_{11}^{5/4}\rho_{-3}^{7/8}d_{28}^2 ,&He ;\cr
2.1\times 10^{-12}\L52^{11/8}\theta_{-1}^{3/2}/r_{10}^{5/4}\rho_{0}^{7/8}d_{28}^2 ,& CO.\cr }
\quad\ergcmsqs~. \label{eq:f_0}
\enq
The rise time of the pulse, and hence the time at which the flux peaks, is
\beq
t_{peak}^{obs} \approx dt_{ang}^{obs}\approx\theta^2 r/2c=\cases{
 5.2\times 10^{-1} \theta_{-1}^2 r_{12.5} , & H;\cr
 1.7\times 10^{-2} \theta_{-1}^2 r_{11} , & He;\cr
 1.7\times 10^{-3} \theta_{-1}^2 r_{10} , & CO,\cr }
{\rm s}, \label{eq:t_peak}
\enq
and the peak flux is
\beq
f_{peak}= \cases{4.4\times
10^{-9}\L52^{5/4}\theta_{-1}/r_{12.5}\rho_{-7}^{3/4} ,&H ;\cr
1.4\times 10^{-10}\L52^{5/4}\theta_{-1}/r_{11}\rho_{-3}^{3/4} ,&He ;\cr
7.4\times 10^{-12}\L52^{5/4}\theta_{-1}/r_{10}\rho_{0}^{3/4} ,&CO .\cr }
\quad\ergcmsqs~. \label{eq:f_peak}
\enq

\section{Cork dynamics following break-out}
\label{sec:dynamics}

At the time $t=t_{exp}$ in the frame moving with Lorentz factor
$\Gamma_h$, the rarefaction wave arrives at the backward shock
decelerating the jet; the pressure behind this shock drops, and it
can no longer decelerate the fast jet plasma. A new system of
forward-backward shocks is formed, which propagates into the
expanding cork plasma (see figure \ref{fig:shocks} for a schematic
description). As the second forward shock breaks through the
expanding cork, a second X-ray flash is produced. We consider the
cork dynamics following the jet emergence in the relativistic
limit, an approximation which is increasingly well satisfied.

\subsection{Second shock}
\label{sec:2ndshock}

In the original frame of the shocked cork, moving with $\Gamma_h$,
the shocked plasma expands, following the penetration of the
rarefaction wave, with velocity comparable to the speed of sound,
$\approx c/\sqrt{3}$. Adding (relativistically) this velocity to
$\Gamma_h$, we find that the cork expands, in the source frame,
with $\Gamma'_h\approx(\sqrt{3/2}+1/\sqrt{2})\Gamma_h= 2\Gamma_h$.
Due to the expansion, the proper density of the expanding cork
drops as $\rho=\rho_0 x^{-3}=4\Gamma_h\rho x^{-3}$. Let us denote
by $\Gamma_{h2}$ the Lorentz factor (in the source frame) to which
the rarefied cork plasma is accelerated by the second shock driven
by the jet. As the cork density decreases with time, the second
shock accelerates. Balancing the pressure behind the new forward
shock, $3P_{h2}=4(\Gamma_{h2}/2\Gamma'_h)^2 4\Gamma_h\rho x^{-3}
c^2= 4\Gamma_{h2}^2\rho c^2/4\Gamma_h x^{3}$ (assuming the shock
is relativistic), with the pressure behind the new backward shock,
$3P_r=4(\Gamma_j/2\Gamma_{h2})^2 n_j m_p c^2$, we find that
$\Gamma_{h2}$ is given by an expression similar to that determining
$\Gamma_h$, with $\rho$ replaced by $\rho/4\Gamma_h x^{3}$. Using
Eq. (\ref{eq:Gammah}) we therefore find
\begin{equation}
\frac{\Gamma_{h2}}{\Gamma_h}=(4\Gamma_h x^3)^{1/4}.
\label{eq:Gammah2}
\end{equation}
Note that if we had assumed that the new forward shock driven into
the cork is not relativistic, pressure balance gives
$\beta_{h2}=x^{3/2}\Gamma_h^{1/2}/2$ for the shock velocity (in the
frame where cork plasma is at rest), which implies that the shock
is indeed relativistic.

In the frame moving with $\Gamma_h$, the forward shock Lorentz
factor is $\Gamma_s=\sqrt{2}\Gamma_{h2}/2\Gamma_h=(\Gamma_h
x^3)^{1/4}$. The shock breaks through the cork at time $t$ for
which the distance it travels, given by $\int_{t_{exp}}^t dt
(1-1/2\Gamma_s^2)c$, equals the thickness of the expanded cork,
$x(t)\Delta_0$. For $\Gamma_h\rightarrow\infty$,
$\Gamma_s\rightarrow\infty$ and second shock break out occurs when
$x\Delta_0=c(t-t_{exp})=\sqrt{3}\Delta_0(x-2)$, i.e. for
$x=2\sqrt{3}/(\sqrt{3}-1)=4.7$. For finite $\Gamma_h$, crossing
occurs at larger $x$, e.g. at $x=5$ for $\Gamma_h=4$. We therefore
adopt $x=5$ for the crossing time. Note, that in this estimate we
have assumed that $\Gamma_{h2}$ is given by Eq. (\ref{eq:Gammah2})
throughout the second crossing. However, the cork is composed of
shocked stellar plasma, and shocked jet plasma.
Eq. (\ref{eq:Gammah2}) holds for propagation through the shocked
stellar plasma, while shock propagation in the shocked jet plasma
is faster, due to its lower density. However, since the shock
crossing time for $\Gamma_h=4$ ($x=5$) is close to that obtained
for $\Gamma_s\rightarrow\infty$, we neglect the small correction
implied for the crossing time.

Following the second shock crossing, the new (proper) thickness of
the shocked cork plasma is $\Delta_2\approx
x\Delta_0/[4(\Gamma_{h2}/2\Gamma'_h)]=(x_c/4\Gamma_h)^{1/4}\Delta_0$,
where $x_c\approx 5$ is the value of the expansion factor at shock
break-out.  The resulting second X-ray flash properties are estimated in
\S\ref{sec:2nd3dpulses}, following the same reasoning of \S\ref{sec:emergence}.
We give here only the estimates for the dynamical cork parameters
following the second crossing:
\begin{equation}
\frac{t_{exp2}}{t_{exp}}=\frac{\Delta_2}{\Delta_0}=
\left(\frac{x_c}{4\Gamma_h}\right)^{1/4},
\label{eq:cross2}
\end{equation}
and
\begin{equation}
t_{exp2}^{obs}\equiv\frac{t_{exp2}}{2\Gamma_{h2}}=
\frac{t_{exp}^{obs}}{2\sqrt{x_c\Gamma_h}}=\cases{
4\times 10^{-1}\L52^{-3/8}\theta_{-1}r_{12.5}^{7/4}\rho_{-7}^{3/8},& H;\cr
3\times 10^{-2}\L52^{-3/8}\theta_{-1}r_{11}^{7/4}\rho_{-3}^{3/8},& He;\cr
8\times 10^{-3}\L52^{-3/8}\theta_{-1}r_{10}^{7/4}\rho_{0}^{3/8},& CO.\cr}
{\rm s}.  \label{eq:texp2}
\end{equation}
The proper thickness $\Delta_2$ given by Eq. (\ref{eq:cross2}) is
determined by the difference between the propagation speed of the
(second) shock and the speed of the fluid behind the shock. This
thickness, which is essentially the product of the speed
difference and the shock life time, therefore grows with shock
life time. The proper thickness saturates, i.e. ceases to
increase, when the mass flux across the shock equals the tangential
mass flux (which is proportional to the thickness). In
\S\ref{sec:jet-prop}, we have determined the proper thickness
$\Delta_0$ of the shocked cork plasma by balancing the two mass
fluxes. Using the same argument here would have given
$\Delta_2=\Delta_0$. However, Eq. (\ref{eq:cross2}) gives
$\Delta_2<\Delta_0$, which implies that the $\Delta_2$ does not
reach its asymptotic value during shock crossing.

\begin{figure}[htb]
\centering \epsfig{figure=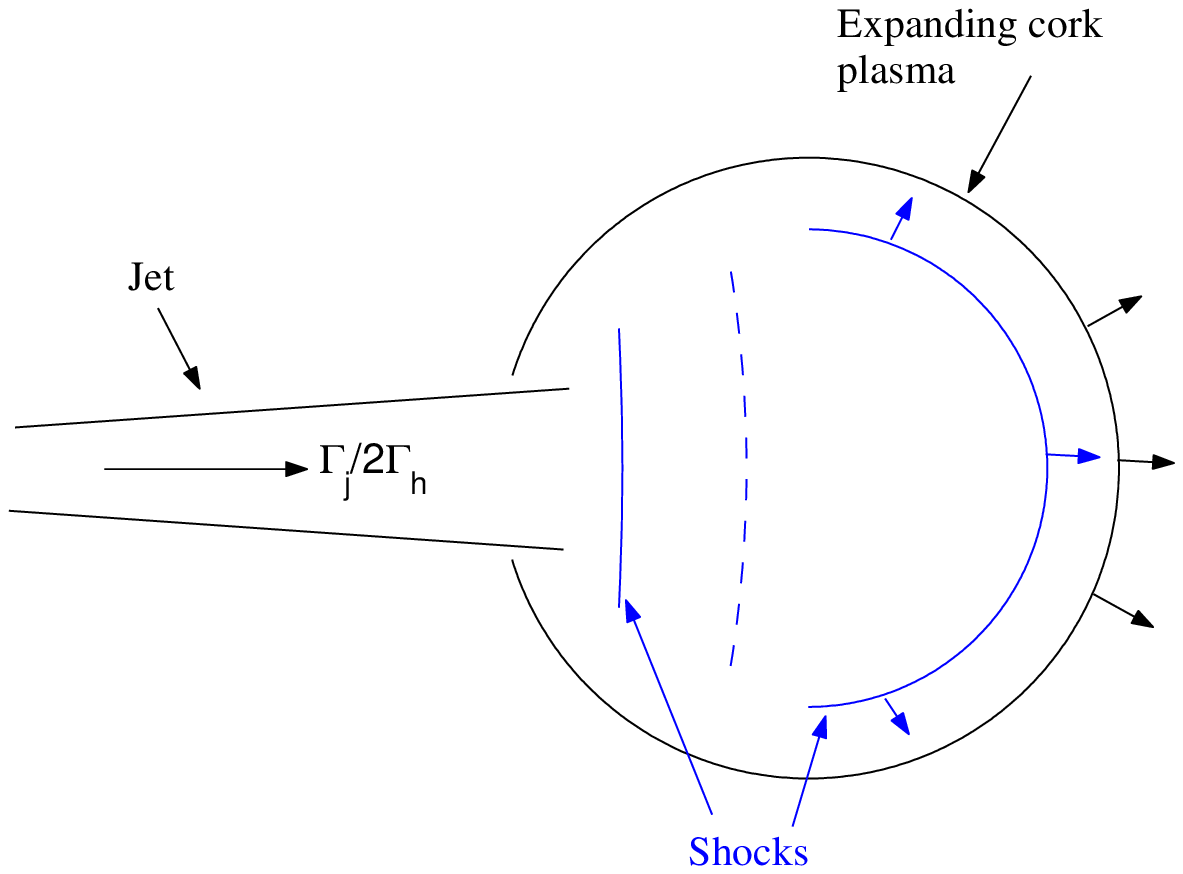} \caption{A schematic
description of the flow during second shock propagation through
the cork, as seen in the "isotropic cork frame" moving with
$\Gamma_h$.}
   \label{fig:cork_struct}
\end{figure}

Plasma expansion in the frame moving with $\Gamma_{h2}$ is not
isotropic. Following the first shock break-out, the plasma expands
roughly isotropically in the frame moving with $\Gamma_h$, to
which we refer in what follows as the "isotropic cork frame". As
the second shock propagates through this expanding plasma, the
flow ahead of the shock is not uniform, and the geometry of the
shock front is influenced by the shape of fluid streamlines ahead
of the shock: the shock will no longer propagate radially in the
source frame. This situation is schematically described in figure
\ref{fig:cork_struct}. Since the Lorentz factor of the second
shock, with respect to the fluid ahead of the shock, is not very
large, see Eq. (\ref{eq:Gammah2}), we expect a large modification
of the shock front geometry due to the non-uniform flow ahead of
the shock. Assuming uniform pressure behind the shock implies
uniform velocity of the shock front with respect to the fluid
velocity ahead of the shock, i.e. radial shock expansion in the
isotropic cork frame (moving with $\Gamma_h$), where the pre-shock
fluid expands isotropically. Following the second shock crossing,
therefore, the shocked cork plasma fills, in the isotropic cork
frame, a semi-spherical shell of proper thickness $\Delta_2$. This
shell expands radially in the isotropic cork frame, with a Lorentz
factor $\Gamma_{h2}/2\Gamma_h$. The fluid propagating along the
jet axis propagates in the source frame with Lorentz factor
$\Gamma_{h2}$, and expands over an observed expansion time given
by Eq. (\ref{eq:texp2}).

\subsection{Third shock}
\label{sec:3dshock}

Following the second shock break-out, a second rarefaction wave
propagates into the newly shocked cork, following which a third
shock crosses the rarefied cork. We denote by $\Gamma_{h3}$ the
Lorentz factor in the source frame to which the rarefied cork
plasma along the jet axis is accelerated by the third shock. Here
too, we expect the third shock to propagate radially, with Lorentz
factor $\Gamma_{h3}/2\Gamma_h$, in the isotropic cork frame. We
define again the expansion factor as $x\equiv 1+c_s t/\Delta_2$,
where $t$ is the proper time, measured in the fluid rest frame
from second shock break-out. Time $\tilde{t}$ measured in the
isotropic cork frame is related to the proper time through
$\tilde{t}=(\Gamma_{h2}/2\Gamma_h)t$. The radius of the expanding
shell in the isotropic cork frame at the second shock break-out is
$R\approx x_c c_s t_{exp}=x_c\Delta_0\approx2\theta r$. This
radius grows with time to $R+\tilde{t}c\approx R+(\Gamma_h
x_c^3)^{1/4}ct/\sqrt{2}=R[1+\sqrt{3/2}(\Gamma_h/x_c)^{1/4}t/t_{exp}]
=R[1+(\sqrt{3}/2)t/t_{exp2}]\approx x R$. Thus, the proper density
of the expanding cork is $\rho=\rho_1 x^{-3}=
4(\Gamma_{h2}/2\Gamma'_h)4\Gamma_h\rho (x x_c)^{-3}=
4\Gamma_{h2}\rho(x x_c)^{-3}$. Balancing again the pressure behind
the new forward shock, $3P_{h3}=4(\Gamma_{h3}/2\Gamma'_{h2})^2
4\Gamma_{h2}\rho (x x_c)^{-3} c^2= 4\Gamma_{h3}^2\rho
c^2/4\Gamma_{h2} (x x_c)^{3}$ (assuming a relativistic shock),
with the pressure behind the new backward shock,
$3P_r=4(\Gamma_j/2\Gamma_{h3})^2 n_j m_p c^2$, we find that
$\Gamma_{h3}$ is given by an expression similar to that
determining $\Gamma_h$, with $\rho$ replaced by $\rho/4\Gamma_{h2}
(x x_c)^{3}$. Using Eq. (\ref{eq:Gammah}) we therefore find
\begin{equation}
\frac{\Gamma_{h3}}{\Gamma_h}=(4\Gamma_{h2} x_c^3 x^3)^{1/4}
=2^{5/8}x_c^{15/16}x^{3/4}\Gamma_h^{5/16}.
\label{eq:Gammah3}
\end{equation}
If we had assumed that the new forward shock is not relativistic,
pressure balance would give $\beta_{h3}=0.5x_c^{3/8}x^{3/2}\Gamma_h^{1/8}$,
which implies that the shock is relativistic. Arguments similar to
those presented above imply that the third shock crossing also
occurs at $x=x_c\approx5$. At the end of the third shock crossing
we therefore have
\begin{equation}
\frac{t_{exp3}}{t_{exp}}=\frac{\Delta_3}{\Delta_0}=
\left(\frac{x_c}{4\Gamma_h}\right)^{5/16}, \label{eq:cross3}
\end{equation}
and
\begin{equation}
t_{exp3}^{obs}\equiv\frac{t_{exp3}}{2\Gamma_{h3}}=
\frac{t_{exp}^{obs}}{x_c^{11/8}(4\Gamma_h)^{5/8}}=\cases{
7.5\times 10^{-2}\L52^{-13/32}\theta_{-1}r_{12.5}^{29/16}\rho_{-7}^{13/32},&H;\cr
7.0\times 10^{-3}\L52^{-13/32}\theta_{-1}r_{11}^{29/16}\rho_{-3}^{13/32},&He;\cr
1.7\times 10^{-3}\L52^{-13/32}\theta_{-1}r_{10}^{29/16}\rho_{0}^{13/32},&CO.\cr}
{\rm s} . \label{eq:texp3}
\end{equation}
Finally, we note that at the end of the third shock crossing, the
radius of the expanded cork, in the isotropic cork frame, is $x_c
R=x_c^2\Delta_0$, and the distance $\delta r$ propagated by the
cork along the jet axis is comparable to the stellar radius,
$\delta r\approx c \Gamma_{h2}x_c t_{exp2}\approx
\sqrt{3}x_c^2\Gamma_h\Delta_0\approx\Gamma_h\theta_{-1} r$.


Following the third rarefaction wave, the plasma is accelerated to
$\Gamma'_{h3}\approx2\Gamma_{h3}\approx 50\Gamma_h^{21/16}$, for
which $\Gamma_j/2\Gamma'_{h3}\approx\Gamma_j/100\Gamma_h^{21/16}$.
Thus, following the third shock crossing the remnants of cork
plasma in the jet path are accelerated to a high Lorentz factor.
A strong fourth shock wave will be driven by the jet into this
plasma only for $\Gamma_j\gg100$.

\section{X- and $\gamma$-ray pulses from the 2nd and 3d shocks}
\label{sec:2nd3dpulses}

\subsection{Thermal pulses}
\label{sec:thermal}

The thermal emission following the second and third shock
crossings can be estimated using arguments similar to those of
\S\ref{sec:emergence}. The energy density in the shocked cork
plasma is proportional to $L/r^2\Gamma_{hi}^2$, where subscripts
$i=2,3$ stand for second and third shock crossings respectively.
Thus, $T_2=T_r(\Gamma_{h2}/\Gamma_h)^{-1/2}$ and
$T_3=T_r(\Gamma_{h3}/\Gamma_h)^{-1/2}$, where $T_r$ is given by
Eq. (\ref{eq:Tr}). Using Eqs. (\ref{eq:Gammah2}) and
(\ref{eq:Gammah3}) we obtain for the temperature of the observed
pulses
\beq
\Gamma_{h2}T_2=(4\Gamma_h)^{1/8}x_c^{3/8}\Gamma_h T_r= \cases{
 6.2 {\L52^{13/32}}/{r_{12.5}^{13/16} \rho_{-7}^{5/32}},&H;\cr
 2.3\times 10^1 {\L52^{13/32}}/{r_{11}^{13/16} \rho_{-3}^{5/32}},&He;\cr
 5.1\times 10^1 {\L52^{13/32}}/{r_{10}^{13/16} \rho_{0}^{5/32}},&CO,\cr}
{\rm keV}, \label{eq:T1}
\enq
\beq
\Gamma_{h3}T_3=2^{5/16}x_c^{27/32}\Gamma_h^{5/32}\Gamma_h T_r=\cases{
 1.4\times 10^1{\L52^{53/128}}/{r_{12.5}^{53/64}\rho_{-7}^{21/128}},&H;\cr
 5.2\times 10^1{\L52^{53/128}}/{r_{11}^{53/64}\rho_{-3}^{21/128}},&He;\cr
 1.1\times 10^2{\L52^{53/128}}/{r_{10}^{53/64}\rho_{0}^{21/128}},&CO.\cr}
{\rm keV}. \label{eq:T2}
\enq

Radiation emitted following the second shock crossing arrives at a
distant observer lying along the jet axis from a cone around the
jet axis of opening angle
$1/(\Gamma_{h2}/2\Gamma_h)=2\Gamma_h/\Gamma_{h2}$ in the isotropic
cork frame, where the shell radius is $R=x_c\Delta_0$ .
The angular time spread of the pulse is therefore given by
$dt_{ang2}^{obs}=(2\Gamma_h/\Gamma_{h2})^2R/2(2\Gamma_h)c=\Gamma_h
x_c\Delta_0/\Gamma_{h2}^2c$. Comparing this time to the observed
expansion time, $t_{exp2}^{obs}=\sqrt{3}\Delta_2/2\Gamma_{h2}c
=\sqrt{3}x_c\Gamma_h\Delta_0/2\Gamma_{h2}^2c$, we find
$dt_{ang2}^{obs}/t_{exp2}^{obs}=2/\sqrt{3}\approx1$. Thus, the
second pulse is ``smeared'' over a duration $dt_{ang2}^{obs}$.
Note that, had we assumed radial expansion in the source frame, the
angular time spread would have been
$r/2\Gamma_{h2}^2c=(r/2x_c\Gamma_h\Delta_0)dt_{ang2}^{obs}
\approx(2.5/\Gamma_h\theta_{-1})dt_{ang2}^{obs}$. Thus, the
resulting pulse duration is not sensitive to the detailed geometry
of the shock front.

The proper energy in the cork plasma following the second shock
crossing is $E_2=2\pi R^2\Delta_2(aT_2^4)= 2\pi(x_c\Delta_0)^2
(\Gamma_{h}x_c\Delta_0/ \Gamma_{h2})
a(\Gamma_{h2}/\Gamma_h)^{-2}T_r^4=2\pi\Delta_0^3
(\Gamma_{h}x_c/\Gamma_{h2})^3aT_r^4\approx
0.3(\Gamma_{h}x_c/\Gamma_{h2})^3E_0$. Since the optical depth
following the first crossing is reduced,
$\tau_{T,s2}=x_c^{-2}\tau_{T,s}$, the flux of the second pulse is
$f_2\approx\Gamma_{h2}(3E_2/\sqrt{2\tau_{T,s2}})/2t_{exp2}^{obs}
\pi (d/\Gamma_h^2)\approx 0.3x_c^3 f_0$,
 \beq
f_2\approx0.3x_c^{3}f_0=\cases{
8.6\times 10^{-8}
{\L52^{11/8}\theta_{-1}^{3/2}}/{r_{12.5}^{5/4}\rho_{-7}d_{28}^{7/8}},& H;\cr
6\times 10^{-10}
{\L52^{11/8}\theta_{-1}^{3/2}}/{r_{11}^{5/4}\rho_{-3}d_{28}^{7/8}},& He;\cr
1.1\times 10^{-11}
{\L52^{11/8}\theta_{-1}^{3/2}}/{r_{10}^{5/4}\rho_{0}d_{28}^{7/8}},& CO. \cr}
\ergcmsqs~. \label{eq:f_2}
\enq
The duration of this pulse is given by Eq. (\ref{eq:texp2}).

Radiation emitted following the third shock crossing arrives at a
distant observer lying along the jet axis from a cone around the
jet axis of opening angle
$1/(\Gamma_{h3}/2\Gamma_h)=2\Gamma_h/\Gamma_{h3}$ in the isotropic
cork frame, where the shell radius is $R=x_c^2\Delta_0$ . The
angular time spread of the pulse is therefore given by
$dt_{ang3}^{obs}=(2\Gamma_h/\Gamma_{h3})^2R/2(2\Gamma_h)c=\Gamma_h
x_c^2\Delta_0/\Gamma_{h3}^2c$. Comparing this time to the observed
expansion time, $t_{exp3}^{obs}=\sqrt{3}\Delta_3/2\Gamma_{h3}c
=\sqrt{3}x_c^2\Gamma_h\Delta_0/2\Gamma_{h3}^2c$, we find
$dt_{ang3}^{obs}/t_{exp3}^{obs}=2/\sqrt{3}\approx1$. Thus, the
third pulse is ``smeared'' over a duration $dt_{ang3}^{obs}$. If
we had we assumed radial expansion in the source frame, the
angular time spread would have been
$r/2\Gamma_{h3}^2c=(r/2x_c^2\Gamma_h\Delta_0)dt_{ang3}^{obs}
\approx(2.5/x_c\Gamma_h\theta_{-1})dt_{ang3}^{obs}$. Thus, the
resulting pulse duration is again not sensitive to the detailed
geometry of the shock front.

The proper energy in the cork plasma following the third shock
crossing is $E_3=2\pi
R^2\Delta_3(aT_3^4)=2\pi(x_c^2\Delta_0)^2(\Gamma_{h}x_c^2\Delta_0/
\Gamma_{h3})a(\Gamma_{h3}/\Gamma_h)^{-2}T_r^4=2\pi\Delta_0^3
(\Gamma_{h}x_c^2/\Gamma_{h3})^3aT_r^4\approx
0.3(\Gamma_{h}x_c^2/\Gamma_{h3})^3E_0$. Since the optical depth
following the third crossing is reduced,
$\tau_{T,s3}=x_c^{-4}\tau_{T,s}$, the flux of the third pulse is
$f_3\approx\Gamma_{h3}(3E_3/\sqrt{2\tau_{T,s3}})/2t_{exp3}^{obs}
\pi (d/\Gamma_h^2)\approx 0.3 x_c^6 f_0$,
 \beq
f_3\approx0.3x_c^6 f_0=\cases{ 1.1\times 10^{-5}
{\L52^{11/8}\theta_{-1}^{3/2}}/{r_{12.5}^{5/4}\rho_{-7}d_{28}^{7/8}},&
H;\cr 7.7\times 10^{-8}
{\L52^{11/8}\theta_{-1}^{3/2}}/{r_{11}^{5/4}\rho_{-3}d_{28}^{7/8}},&
He;\cr 1.4\times 10^{-9}
{\L52^{11/8}\theta_{-1}^{3/2}}/{r_{10}^{5/4}\rho_{0}d_{28}^{7/8}},&
CO. \cr} \ergcmsqs~. \label{eq:f_3} \enq The X-ray pulses from the
three shocks are schematically shown in Figure
\ref{fig:xr_precursor}, appearing as precursors to the main burst
emission.

\subsection{Non-thermal emission}
\label{sec:non-thermal}

As the shock approaches the shell edge, where the optical depth
ahead of the shock drops to $\sim1$, the radiation will escape and
the shock will no longer be radiation dominated. It will most
likely become collisionless. Waxman \& Loeb (2001) have shown
that, for a mildly relativistic shock, the ratio of
electro-magnetic instability growth rate and ion-ion (Coulomb)
collision rate is $\nu_{\rm EM}/\nu_{ii}\approx
10^7\rho_{-10}^{1/2}$. The density at the third shock crossing is
smaller than the initial stellar density by a factor $\sim
x_c^{-6}$. Thus, for all the progenitors considered here, we have
$\nu_{\rm EM}/\nu_{ii}\gg1$, and the shock is likely to become
collisionless, rather than collisional, as it approaches the edge
of the cork, where the optical depth becomes of order 1. If a
fourth shock crossing takes place, i.e. if $\Gamma_j\gg100$, the
fourth shock will be collisionless as well. For the BSG
progenitor, the optical depth of the cork shell during third shock
crossing is reduced to $\sim10$. In this case, therefore, a
significant fraction of the internal energy of the cork may be
emitted as radiation from electrons accelerated to a non-thermal
distribution by the collisionless shock.

We estimate below the non-thermal emission from the third shock as
it becomes collisionless. In order to derive an estimate that is
(nearly) independent of the details of the preceding cork shocks
dynamics, we use the following line of arguments. The
isotropic-equivalent mass of plasma shocked by the collisionless
shock is $M_{CL}\approx4\pi(\Gamma_h\theta_{-1} r)^2\tau
m_p/\sigma_T$, where we have taken $r_{CL}=\Gamma_h\theta_{-1} r$
as the shock radius (due to expansion of the plasma during the 3rd
shock crossing, see \S\ref{sec:dynamics}) and $\tau\sim1$ is the
optical depth ahead of the shock at the point it becomes
collisionless ($\tau m_p/\sigma_T$ is the corresponding column
density). We have shown that the Lorentz factors of the shocks, in
the frame at which the plasma ahead of the shocks is at rest, are
of order a few. Thus, the isotropic-equivalent thermal energy
generated by the collisionless shock is
\beq E_{CL}\approx \Gamma_j\Gamma_i M_{CL} c^2 =\cases{
7.7\times10^{50} \tau\Gamma_i\Gamma_{j,2.5}\L52^{1/2}\theta_{-1}^2
  r_{12.5}\rho_{-7}^{-1/2},& H;\cr
2.6\times10^{47} \tau\Gamma_i\Gamma_{j,2.5}\L52^{1/2}\theta_{-1}^2
  r_{11}\rho_{-3}^{-1/2},& He;\cr
8.6\times10^{44} \tau\Gamma_i\Gamma_{j,2.5}\L52^{1/2}\theta_{-1}^2
  r_{10}\rho_{0}^{-1/2},& CO,\cr} {\rm erg},
\label{eq:E_CL}
\enq
where $\Gamma_i\sim$ a few is the collisionless shock Lorentz
factor (in the plasma frame). We have shown in
\S\ref{sec:dynamics} that the proper thickness of the shocked cork
shell following each shock passage is close to $\Delta_0$. The
proper energy density behind the collisionless shock is therefore
$u_{CL}\approx (E_{CL}/\Gamma_j)/4\pi
r_{CL}^2(\Delta_0\tau/\tau_{CL})$, where $\tau_{CL}\sim10$ is the
total optical depth of the shell at the stage where the
collisionless shock is formed. Assuming that a fraction $\xi_e$
($\xi_B$) of the thermal energy is carried by accelerated
electrons (magnetic field), the characteristic frequency of
synchrotron photons emitted by the electrons, accelerated to a
characteristic Lorentz factor $\gamma_e\approx\xi_e\Gamma_i(m_p/m_e)$, is
\beq
h\nu_{syn} \approx \cases{
7.9 \Gamma_i^{5/2}\xi_e^2
 \sqrt{\tau_{CL}\xi_B}\Gamma_{j,2.5}\theta_{-1}^{-1/2}r_{12.5}^{-1/2},&H;\cr
4.3\times 10^1 \Gamma_i^{5/2}\xi_e^2
 \sqrt{\tau_{CL}\xi_B}\Gamma_{j,2.5}\theta_{-1}^{-1/2}r_{11}^{-1/2},&He;\cr
1.4\times 10^2 \Gamma_i^{5/2}\xi_e^2
 \sqrt{\tau_{CL}\xi_B}\Gamma_{j,2.5}\theta_{-1}^{-1/2}r_{10}^{-1/2},&CO.\cr}
\label{eq:nusy}{\rm MeV}.
\enq
Thus, the third thermal (hard) X-ray pulse is accompanied by a non-thermal
MeV gamma-ray pulse, with similar energy flux and duration.

An IC component is also expected, due to the thermal X-ray photons
emitted from the cork which are up-scattered by the fast jet,
$\Gamma_j\sim10^2$, as it emerges from the cork (Ramirez-Ruiz, et
al. 2002). The fluence of this IC component is significantly lower
than the value estimated in the above reference, which
was based on a simplified treatment assuming that all the thermal
X-ray photons in the cork are up-scattered by the relativistic jet.
However, as shown in \S\ref{sec:1stflash}, only a
fraction $\tau_{T,0}^{-1/2}\ll1$ of the thermal X-ray photons escape
the cork and are available for up-scattering [see Eq.~(\ref{eq:Etot})];
the value of $\tau_{T,0}^{-1/2}$ ranges from $10^{-2}$ for the BSG
progenitor to $10^{-4}$ for the CO progenitor [see Eq.~(\ref{eq:tauTs})].
Also, as shown in \S\ref{sec:dynamics},  the fast jet emerges from the
cork only after third shock crossing, at a time (measured in the star's
frame) $\approx\Gamma_{h2}x_c t_{exp2}\approx x_c^2\Delta_0/c$ after
the first cork break-out. This time is much longer than the transverse
light crossing time of the jet, $\simeq\Delta_0/c$. Thus, most of the
X-ray photons escape out of the jet's path prior to the emergence of
the jet from the cork. We therefore conclude that the non-thermal
emission is dominated by the collisionless emission given by
Eqs.~(\ref{eq:E_CL}) and (\ref{eq:nusy}).

A second IC component may be produced if there is a $\Gamma \sim
10\Gamma_1$ fore-runner shell ejected ahead of the main cork, as discussed in
\S \ref{sec:emergence-dynamics}, from the up-scattering of stellar
photons by the fore-runner. The number density of stellar photons
drops as $r^{-2}$, so most of the fluence is produced by scattering
of stellar photons near $R_\ast$, where $R_\ast$ is the stellar
radius. The (isotropic equivalent) energy of this IC component is
$\sim L_\ast (R_\ast/c)\Gamma^2 \sim 10^{37} \Gamma_1^2 (L_\ast/L_\odot)$
erg, where $L_\ast$ is the stellar luminosity. This is much smaller than
other high energy components, such as equation (\ref{eq:E_CL}).

\section{Discussion}
\label{sec:disc}

We have shown that the emergence of a jet in a collapsar model of GRB
leads to a series of successive, increasingly shorter and harder thermal
X-ray pulses. These are caused by successive cycles of shock waves and
rarefaction waves reaching the outer and inner ends of the ``cork" of
stellar material being pushed ahead of the jet as it passes beyond the
boundary of the stellar envelope. The Lorentz factor of the main  bulk
of the cork increases with each succeeding and increasingly relativistic
shock that goes through it. A small fraction of the cork mass may
be accelerated to larger Lorentz factors and run ahead of the cork,
without changing significantly the development of the thermal pulses
from the main bulk of the cork.

The expansion and acceleration of the main bulk of the cork lead
to a decrease of the cork scattering optical depth, photon
diffusion times and expansion times in the observer frame for the
successive shocks. As a result the successive shocks have
increasingly higher  bolometric fluences and shorter durations, as
well as harder peak photon energies. However, for observations in
a fixed low energy X-ray band, e.g 2 keV, the specific fluence in
erg/cm$^2$/keV would be reduced by a Rayleigh-Jeans factor
$(E_{obs}/E_{pk})^2$ and would appear almost constant or slowly
declining as the pulse number increases. On the other hand, in a
hard X-ray detector, e.g. $\simg 20$ keV, the successive pulses
would show a steeply increasing fluence.

The results discussed here are based on analytical 1-D
calculations in the relativistic limit. As argued in \S~2 and in
\S~3, the approximations inherent in such an analytical treatment
are reasonable for the range of parameters considered, but the
flows considered are quite complex, and definitive results must
await confirmation from more accurate numerical simulations.
Current 2-D relativistic collapsar jet simulations (e.g. Aloy
\etal 2000, Zhang \etal, 2002) do not so far have the dynamic
range and resolution necessary to distinguish multiple shocks and
rarefaction waves upon emergence of the jet, but provide valuable
informations about the overall jet dynamics. The approximate
analytical treatment presented here fulfill in the meantime an
exploratory role, and provide insights both for future numerical
and observational developments.

The temporal structure discussed provides distinct features, which should
help to identify the phenomenon.
We have argued (\S \ref{sec:jet-prop}) that since jets are likely
to accelerate to significantly relativistic velocities as they
enter a much lower density H-envelope, the crossing time for a blue
supergiant envelope may be only slightly longer than the core crossing
time $t_{core}\sim 30r_{11}$ s for He or CO dwarf stars.
The initial rise-time of the first pulse
leading to the peak flux, given by the angular smearing time, is
$t_{pk}^{obs}\sim [500,~20,~2]$ ms for a blue supergiant (H) star, a He
star and a CO star, occurring while the cork has not yet had time to expand.
This is followed by a decay $f\propto t^{-1/2}$ lasting an expansion time
$t_{exp}^{obs}\sim [2,~0.15,~0.025]$ s for the [H, He, Co] cases.
Thereafter the flux drops as $f\propto t^{-2}$. The subsequent pulses
have similar rise and decay times, given approximately by (increasingly
shorter) observer-frame expansion times.

For stars of greater compactness (smaller radius) the characteristic
bolometric fluxes, timescales and fluences of each pulse are lower and
shorter while the peak temperatures are higher. The typical fluences
from a Hubble distance $d\sim 10^{28}$ cm range from ${\cal F} \sim
[5\times 10^{-9},7\times 10^{-12},5\times 10^{-14}]$  erg/cm$^2$
for the first pulse, to ${\cal F} \sim
[8\times 10^{-7},5\times 10^{-10},2\times 10^{-12}]$ erg/cm$^2$
for the third pulse, while the peak temperatures range from
$T_{pk}^{obs}\sim [1.6,~12,~50]$ keV for the first pulse, to
$T_{pk}^{obs}\sim [14,~52,~110]$ keV for the third pulse in
[H, He, CO] stars respectively.

After the third shock, the cork will have reached Lorentz factors
comparable to the observationally inferred Lorentz factors in bursts
($\Gamma \sim 100)$. This shock may become collisionless, leading
also to an additional non-thermal component of energy and duration
comparable to the thermal third pulse, with characteristic photon
energy $E_\gamma \sim [8,~45,~140]$ MeV for [H, He, CO] stars.

With the fluences calculated here it is apparent that it would be
easier to detect the X-ray pulses from extended GRB progenitors,
such as blue supergiant (H) stars.  The more compact progenitors
(He or CO stars) would have lower fluences and be harder to
detect, except for nearby objects, in which case they would be
distinguishable from blue supergiant progenitors through their shorter
peak durations and harder spectra. The characteristic precursor
thermal pulses, including the third thermal and non-thermal pulses
described here, would precede the conventional MeV range emission
by an appreciable timescale $\sim t_{exp}^{obs} \sim 2.2
\L52^{-1/2}\theta_{-1}^{3/2} r_{12.5}^{3/2}\rho_{-7}^{1/4}$ s for
blue supergiant (H) progenitors, or by shorter timescales $\sim
0.15,~0.025$ s for He or CO progenitors.

It is conventionally assumed that the conditions for the usual successful
GRB to occur is that the central engine powers the jet for timescales
$t_j$ longer than the stellar crossing time $t_{core} \sim 30r_{11}$ s.
However, there may be cases where the jet just breaks through the
outer envelope of the star (whether it be a He, CO or H BSG star),
and does not live much longer afterwards (e.g. $t_j \siml 30$ s).
In this case, electromagnetic detectors will see only the thermal
X-ray pulses and the hard non-thermal gamma-ray pulse ($\simg 10$ MeV, \S
\ref{sec:non-thermal}), which last seconds or less. These would be a new
class of objects, perhaps related to X-ray flash bursts and/or to short
GRBs, especially if such events are more common, and their distance is smaller.

The characteristic timescales as well as the ratios of the amplitudes
and hardnesses of the pulses differ substantially, depending on the
density and extent of the stellar envelope, which provides a potentially
valuable diagnostic for pinning down the conditions in the pre-burst
stellar progenitor.  Instruments to be flown in the next several years,
such as Swift, GLAST, AGILE and others, should be able to detect such
precursor signals in some bursts, and provide valuable clues concerning
the GRB mechanism and their progenitors.

\acknowledgements{This research has been supported by NSF AST0098416,
BSF 9800343, Universities Planning \& Budget Committee, Israel,
and NASA NAG5-9192.}

\begin{figure}[htb]
\centering \epsfig{figure=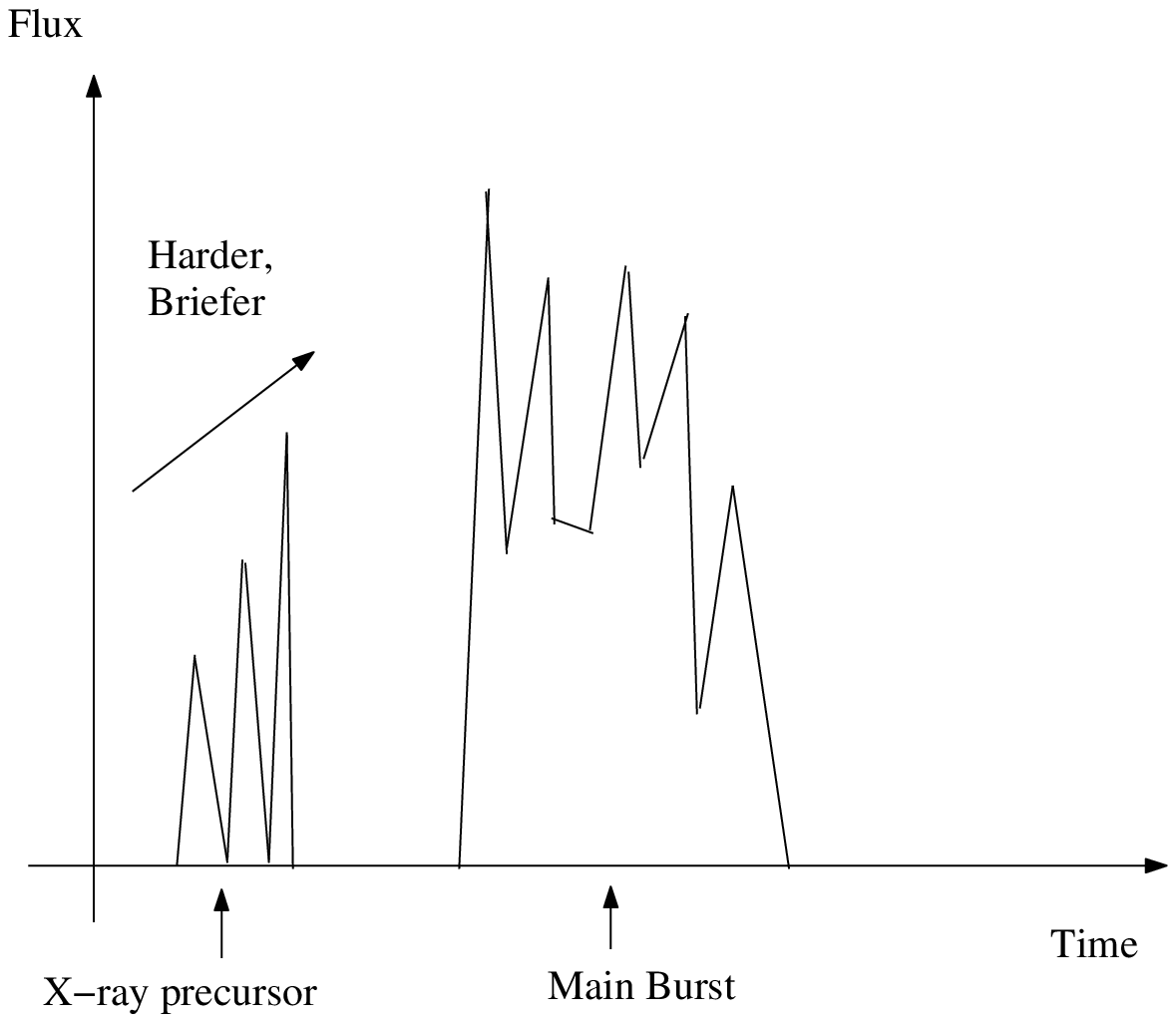, width=4.in,
height=4.in} \caption{ X-ray precursors of increasing flux and
hardness and decreasing duration, resulting from successive shocks
and rarefaction fronts moving across the cork of stellar material
ejected by the GRB jet. }
   \label{fig:xr_precursor}
\end{figure}

\end{document}